\documentclass[twocolumn,showpacs,showkeys,preprintnumbers,amsmath,amssymb]{revtex4}
\usepackage{graphicx}% Include figure files
\usepackage{dcolumn}% Align table columns on decimal point
\usepackage{bm}% bold math

\begin{document}
%\preprint{Submitted to PRL}
\title{The Ramsey method in high-precision mass spectrometry with Penning traps: Experimental results\\}

\author{S.~George$^{1,2}$}
\email[Correspondence to: ]{george@uni-mainz.de}
\author{K.~Blaum$^{1,2}$}
\author{F.~Herfurth$^{1}$}
\author{A.~Herlert$^{3}$}
\author{M.~Kretzschmar$^{2}$}
\author{S.~Nagy$^{2}$}
\author{S.~Schwarz$^{4}$}
\author{L.~Schweikhard$^{5}$}
\author{C.~Yazidjian$^{1}$}

\affiliation{$^{1}$GSI, Planckstra{\ss}e 1, 64291 Darmstadt, Germany} 
\affiliation{$^{2}$Johannes Gutenberg-Universit\"at, Institut f\"ur Physik, 55099 Mainz, Germany} 
\affiliation{$^{3}$CERN, Division EP, 1211 Geneva 23, Switzerland} 
\affiliation{$^{4}$NSCL, Michigan State University, East Lansing, MI 48824-1321, USA}
\affiliation{$^{5}$Ernst-Moritz-Arndt-Universit\"at, Institut f\"ur Physik, 17487 Greifswald, Germany}

\date{\today}

\begin{abstract}
The highest precision in direct mass measurements is obtained with Penning trap mass spectrometry. Most experiments use the interconversion of the magnetron and cyclotron motional modes of the stored ion due to excitation by external radiofrequency-quadrupole fields.
In this work a new excitation scheme, Ramsey's method of time-separated oscillatory fields, has been successfully tested. It has been shown to reduce significantly the uncertainty in the determination of the cyclotron frequency and thus of the ion mass of interest. The theoretical description of the ion motion excited with Ramsey's method in a Penning trap and subsequently the calculation of the resonance line shapes for different excitation times, pulse structures, and detunings of the quadrupole field has been carried out in a quantum mechanical framework and is discussed in detail in the preceding article in this journal by M. Kretzschmar. Here, the new excitation technique has been applied with the ISOLTRAP mass spectrometer at ISOLDE/CERN for mass measurements on stable as
well as short-lived nuclides. The experimental resonances are in agreement with the theoretical predictions and a precision gain close to a factor of four was achieved compared to the use of
the conventional excitation technique.
\end{abstract}
                          
\maketitle

\section{Introduction}
The mass and its inherent connection with the atomic and nuclear binding energy is an important property of a nuclide. Thus, precise mass measurements are eminent for various applications in many fields of physics~\cite{Bla06,Lun03}. The required precision of the atomic mass depends on the physics being investigated. For radionuclides, which often have half-lives considerably less than a second, it ranges from $\delta m/m=10^{-5}$ to below $10^{-8}$ and for stable nuclides even down to $\delta m/m=10^{-11}$. Since the interest in masses of short-lived and stable nuclides arises from a wide range of applications with different requirements on the accuracy, a continuous development of \emph{e.g.} new ion detection and preparation techniques is carried on at several facilities world-wide~\cite{Bol06}.

In 1949 N.F. Ramsey improved the molecular-beam magnetic-resonance method of I.I. Rabi~\cite{Rab38,Rab39} by applying oscillatory fields to the transversing molecular beam in spatially separated regions~\cite{Ram90}. This replacement of the uniformly applied field led to a linewidth reduction to $60\%$ and thus to a more precise determination of the resonance frequency. For a detailed description see~\cite{Ram90} and references therein.

In high-precision mass spectrometry the idea to use time-separated oscillatory fields to manipulate the radial motions of confined ions in a Penning trap was first put forward at ISOLTRAP by G. Bollen and coworkers in 1992~\cite{Bol92} and later tested at the SMILETRAP experiment~\cite{Ber02}. At this time the correct theoretical description of the observed line-shapes was not available. Instead, the observations were discussed qualitatively in terms of the Fourier transform of the applied pulse sequence. However, a study of the precision gain as well as on-line mass measurements with the new excitation scheme were not performed. In the preceding paper~\cite{Kre06} a theoretical description of the obtained lineshape has been developed in a quantum mechanical framework. In the following, the corresponding experimental investigations and the gain in linewidth and precision of the ``Ramsey method'' will be presented.

\section{Theoretical overview}
For a quick orientation we review here in a non-technical manner the physical assumptions forming the basis of the theoretical model, the approximations necessary for analytical calculations, and the results that are most important for mass spectrometry. Details are found in the preceding article~\cite{Kre06}.
\subsection{Penning traps with azimuthal quadrupole excitation}
\subsubsection{The ideal Penning trap}
\begin{figure}
\resizebox{0.4\textwidth}{!}
{ \includegraphics{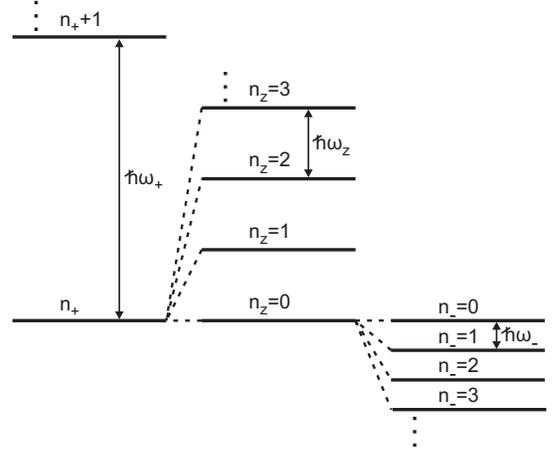} } \vspace*{0.0cm}
\caption{Energy level scheme of the harmonic oscillators for a spin-less     
    charged particle in an ideal Penning trap. $\omega_+$ is the modified cyclotron angular 
    frequency, $\omega_z$ is the angular frequency, and $\omega_-$ is the magnetron 
    angular frequency. $n_+$, $n_z$, and $n_-$ denote the corresponding quantum numbers. The
    total energy is given by the sum of the energies of the three independent harmonic
    oscillators. The contribution of the inverted magnetron oscillator is negative. Zero point
    energies of the oscillators have been subtracted.} 
\label{EnergyLevel}
\end{figure}

The electromagnetic field configuration of an ideal Penning trap consists of a strong homogeneous magnetic field ${\bf B}_{0}=B_{0}{\bf e}_{z}$ in the axial direction and an electrostatic quadrupole field ${\bf E}=-\nabla \Phi_{0}$ derived from the potential\linebreak[4] $\Phi_{0}(x,y,z)=\frac{U}{2z_{0}^{2}+r_{0}^{2}}\cdot(2z^{2}-x^{2}-y^{2})$. The motion of a single ion of mass $m$ and electric charge $q$ in this electromagnetic field is determined by the field parameters $B_0$ and $U$, or equivalently, by the cyclotron frequency $\nu_{c}=\omega_{c}/2\pi=qB_0/(2\pi m)$ and by the axial frequency $\nu_{z}=\omega_{z}/2\pi$ with $\omega_{z}=\sqrt{4qU/(2z_{0}^{2}+r_{0}^{2})}$, where $r_{0}$ and $z_{0}$ denote the inner radius of the ring electrode and the distance of the endcap electrodes from the trap center, respectively. The dynamics of a single ion in an ideal Penning trap is described
by three uncoupled harmonic oscillators: the oscillator of the cyclotron motion with the  frequency $\omega_{+}=\frac{1}{2}(\omega_{c}+\sqrt{\omega_{c}^{2}-2\omega_{z}^{2}})$, the inverted oscillator of the magnetron motion with the frequency
$\omega_{-}=\frac{1}{2}(\omega_{c}-\sqrt{\omega_{c}^{2}-2\omega_{z}^{2}})$,
and the axial oscillator with the frequency $\omega_{z}$. It is useful to introduce the abbreviation $\omega_{1}=\sqrt{\omega_{c}^{2}-2\omega_{z}^{2}}$, so that
$\omega_{\pm}=\frac{1}{2}(\omega_{c}\pm\omega_{1})$. With appropriately chosen canonical coordinates $q_{k}$, $p_{k}$ ($k=+,-,3$) the Hamiltonian for the motion of a single spinless ion can be
written~\cite{Kre06} as
\begin{eqnarray}
H_{0} &=& \omega_{+}\cdot{\textstyle
\frac{1}{2}}(q_{+}^{2}+p_{+}^{2})-\omega_{-}\cdot{\textstyle
\frac{1}{2}}(q_{-}^{2}+p_{-}^{2})\nonumber\\
& &+\omega_{z}\cdot{\textstyle
\frac{1}{2}}(q_{3}^{2}+p_{3}^{2}). \label{ham-q}
\end{eqnarray}
Canonical coordinates offer an easy access to the quantized version of the theory. Annihilation and creation operators for the oscillator quanta are defined by
\begin{eqnarray}
a_{k}&=&\frac{1}{\sqrt{2\hbar}}\left(q_{k}+i\,p_{k}\right) ,\nonumber\\
a_{k}^{\dagger}&=&\frac{1}{\sqrt{2\hbar}}\left(q_{k}-i\,p_{k}\right) ,
\hspace{6mm}(k=+,-,3) \label{a}
\end{eqnarray}
with commutation relations $[a_{\pm},a_{\pm}^{\dagger}]=1$, $[a_{\pm},a_{\mp}^{\dagger}]=0$, and $[a_{\pm},a_{\mp}]=0$. The Hamiltonian then takes the form
\begin{eqnarray}
H_{0} &=& \hbar\omega_{+}(a_{+}^{\dagger}a_{+}+{\textstyle
\frac{1}{2}}) - \hbar\omega_{-}(a_{-}^{\dagger}a_{-}+{\textstyle
\frac{1}{2}}) \nonumber\\
& &+\hbar\omega_{z}(a_{3}^{\dagger}a_{3}+{\textstyle
\frac{1}{2}}) . \label{ham-a}
\end{eqnarray}
The quantized version of the theory provides a clear picture of the energy level scheme associated with the ion motion in an ideal Penning trap (see Fig. \ref{EnergyLevel}).
Therefore, it is valuable for the identification of the interaction with the external rf-quadrupole field that is used to convert the magnetron motion into the cyclotron motion.

Real Penning traps generally possess some anharmonic perturbations that introduce nonlinear terms into the equations of motion and couple the three motional modes, thus preventing us from finding analytical solutions for the ion motion. This problem can be minimized by avoiding large distances of the ion from the trap center. The theoretical discussion assumes also that all anharmonic perturbations and couplings to the axial oscillator mode can be neglected. Therefore the description concentrates exclusively on the cyclotron and magnetron motional modes.

\subsubsection{The ideal Penning trap with azimuthal quadrupole excitation}
\begin{figure}
\centering
{%
 \includegraphics[width=6cm,keepaspectratio]{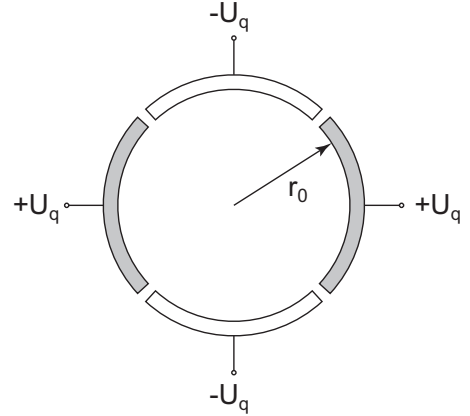}
        % !!! sonst 7 und 8cm
      } \vspace*{0.3cm} 
   \caption{Radial segmentation of the ring electrode (top view) for the application of radiofrequency
    fields. A predominantly quadrupolar field can be generated by applying a radiofrequency between
    pairs of opposing electrodes of the four-fold segmented ring electrode.} \label{Dipol_Quadrupol}
\end{figure}

The application of a radiofrequency potential to the four-fold segmented ring electrode (see Fig.~\ref{Dipol_Quadrupol}) adds to the Hamiltonian (\ref{ham-a}) new terms that are periodic with the driving frequency $\nu_{d}=\omega_{d}/2\pi$ and have a phase $\chi_{d}$. The leading term in a multipolar expansion is the quadrupole contribution $\propto (x^{2}-y^{2})\cos(\omega_{d}t+\chi_{d})$, higher multipoles must be minimized as they introduce non-linearities into the equations of motion. In the quantized version of the theory $x$ and $y$ are components of the position operator of the ion and can be expressed in terms of the creation and annihilation operators of the modified cyclotron and magnetron motional modes. It then becomes apparent that the quadrupole contribution actually describes three physical processes: Most importantly the absorption of an electromagnetic field quantum
$\hbar\omega_{d}$ with simultaneous conversion of a magnetron oscillator quantum $\hbar\omega_{-}$ into a quantum of the cyclotron oscillator $\hbar\omega_{+}$, together with the reverse transition. This process is dominant if the driving frequency $\nu_{d}$ approximately satisfies the resonance condition $\hbar\omega_{d}\approx\hbar\omega_{+}+\hbar\omega_{-}=\hbar\omega_{c}$. It is described by an additional term in the Hamiltonian
\begin{eqnarray}
H_{1}(t)&=& \hbar
g\left(e^{-i(\omega_{d}t+\chi_{d})}a_{+}^{\dagger}(t)a_{-}(t)\right.\nonumber\\
& &\left.+e^{+i(\omega_{d}t+\chi_{d})}
a_{-}^{\dagger}(t)a_{+}(t)\right)\;. \label{quadinteraction}
\end{eqnarray}
The real coupling constant $g$ has the physical dimension of a frequency. It is proportional to the amplitude of the rf-quadrupole field, but its value is also influenced by details of the trap geometry. As shown later, it determines the conversion time $\tau_{c}=\pi/2g$, which is defined as the time required for the full conversion of a state of pure magnetron motion into a state of pure cyclotron motion by a quadrupole field at the resonance frequency $\omega_{c}$ (see final results in Eq.~(\ref{1-pulse})). For each magnetron quantum $\hbar\omega_{-}$ that is annihilated (created) a cyclotron quantum $\hbar\omega_{+}$ is simultaneously created (annihilated). The interaction obviously conserves the total number of quanta $N_{{\rm tot}}=N_{+}+N_{-}$ present in the system and has no simple description in a purely classical picture. The other two physical processes described by the quadrupole contribution are the absorption (emission) of a field quantum $\hbar\omega_{d}$ with simultaneous creation (annihilation) of two quanta of the modified cyclotron oscillator (driving frequency $\omega_{d}\approx 2\omega_{+}$), and the absorption (emission) of a field quantum $\hbar\omega_{d}$ with simultaneous annihilation (creation) of two quanta of the magnetron oscillator (driving frequency $\omega_{d}\approx 2\omega_{-}$)~\cite{Sch93}. These two processes are negligible if $\omega_{d}\approx\omega_{c}$.

The ideal Penning trap with quadrupole excitation is now described by the total Hamiltonian $H(t)=H_{0}+H_{1}(t)$. The resulting Heisenberg equations of motion for the operators $a_{+}(t)$ and $a_{-}(t)$ are linear and time-dependent, they permit a general, exact solution. The equations and their solutions correspond closely to those familiar from the study of magnetic-resonance or the quantum theory of two-level systems. An important result is that the interconversion of the magnetron and cyclotron motional modes by a quadrupole field of frequency $\nu_{d}$ is periodic with the \lq Rabi frequency' $\nu_{R}=\omega_{R}/(2\pi)=\sqrt{(g/\pi)^{2}+(\nu_{d}-\nu_{c})^{2}}$. 

\subsubsection{Ion trajectories}
While the viewpoint of quantum mechanics was very helpful for the identification of the relevant interaction, it is in general sufficient for the application to mass spectrometry to consider the ion motion as a classical motion, \textit{i.e.} following trajectories of macroscopic scale. To work out this aspect expectation values of the quantum mechanical operators are taken with respect to quasiclassical coherent oscillator states. Thus, the annihilation operators $a_{+}(t)$ and $a_{-}(t)$, obtained as solutions of the Heisenberg equations of motion, are translated into two complex functions $\alpha_{+}(t)$ and $\alpha_{-}(t)$ that are denoted as the \lq complex oscillator amplitudes' of the cyclotron and magnetron oscillators at time $t$. The complex conjugate functions $\alpha_{+}^{\ast}(t)$ and $\alpha_{-}^{\ast}(t)$ correspond to the respective creation operators. The explicite solution of the initial value problem is~\cite{Kre06}:
\newpage
\begin{eqnarray}
\alpha_{+}(t) & = & e^{-i(\omega_{+}+\delta/2)t} \left[\left(\cos\frac{\omega_{R}t}{2}+i\frac{\delta}{\omega_{R}}
\sin\frac{\omega_{R}t}{2}\right)\alpha_{+}(0)\right.\nonumber\\
& & \left. -i\frac{2g}{\omega_{R}}
\sin\frac{\omega_{R}t}{2}e^{-i\chi_{d}}\alpha_{-}(0)\right]\; ,\label{alphaplus} \\
\alpha_{-}(t) & = & e^{+i(\omega_{-}+\delta/2)t}\left[
-i\frac{2g}{\omega_{R}}\sin\frac{\omega_{R}t}{2}e^{+i\chi_{d}}\alpha_{+}(0)\right.\nonumber\\
 & & \left.
+\left(\cos\frac{\omega_{R}t}{2}-i\frac{\delta}{\omega_{R}}\sin\frac{\omega_{R}t}{2}
\right) \alpha_{-}(0) \right]  \; \label{alphaminus}.
\end{eqnarray}
In these equations $\delta=\omega_{d}-\omega_{c}$ is the detuning of the driving
quadrupole field, $\omega_{R}=\sqrt{4g^{2}+\delta^{2}}$ the Rabi frequency, $\chi_{d}$ the phase of the quadrupole field at time $t=0$, and $\alpha_{\pm}(0)=|\alpha_{\pm}(0)|\exp[\mp i\chi_{\pm}]$ the initial values of the complex oscillator amplitudes. For example, for an initial state of pure magnetron motion or pure cyclotron motion they are $\alpha_{+}(0)=0$ or $\alpha_{-}(0)=0$, respectively.

\begin{figure}
\resizebox{0.5\textwidth}{!}
{%
 \includegraphics{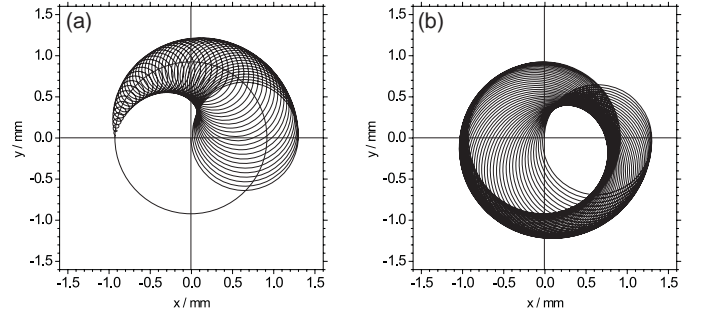}
        % !!! sonst 7 und 8cm
   } \vspace*{0.3cm} 
   \caption{Conversion of pure magnetron motion to pure cyclotron motion by the action
    of an azimuthal quadrupolar field with the cyclotron frequency $\nu_c=\nu_++\nu_-$. Part (a)
    and (b) show the first and second half of the conversion, respectively. The solid circle in
    (a) indicates the initially pure magnetron motion. } \label{ConversionMaCy}
\end{figure}

The practical importance of the complex oscillator amplitudes for our applications lies in their relation to the instantaneous radii for the cyclotron and the magnetron motion,
\begin{equation}
R_{+}(t) \;=\; \sqrt{\frac{2\hbar}{m\omega_{1}}} |\alpha_{+}(t)|
\hspace{5mm},\hspace{5mm} R_{-}(t) \;=\; \sqrt{\frac{2\hbar}{m\omega_{1}}}
|\alpha_{-}(t)| \; .\label{radius}
\end{equation}
Relating the complex oscillator amplitudes to the original Cartesian coordinates and velocities, an explicite parametric representation of the ion trajectories in the $xy$-plane can be derived
\begin{eqnarray}
x(t)+iy(t) & = & e^{-\frac{i}{2}\delta
t}\sqrt{\frac{2\hbar}{m\omega_{1}}}\cdot\left[\left(
\cos\frac{\omega_{R}t}{2}+i\frac{\delta}{\omega_{R}}
\sin\frac{\omega_{R}t}{2}\right)\right.\nonumber\\
 & & \left.\cdot\left(e^{-i\omega_{+}t}\alpha_{+}(0)+e^{-i\omega_{-}t}
\alpha_{-}^{\ast}(0)\right)\right.\nonumber\\
& & \left. -ie^{-i\chi_{d}}\cdot
\frac{2g}{\omega_{R}}\sin\frac{\omega_{R}t}{2}\right.\nonumber\\
& &\left.\left(e^{-i\omega_{+}t}\alpha_{-}(0)
-e^{-i\omega_{-}t}\alpha_{+}^{\ast}(0)\right)\right]\;
.\label{trajectory}
\end{eqnarray}
A graphical representation of an ion trajectory calculated by this approach is shown in  Fig.~\ref{ConversionMaCy}. For a three-dimensional representation one has to add, of course, the
oscillatory axial motion.

\subsection{Excitation schemes}
The precision in the determination of the cyclotron frequency strongly depends on the width of the central peak of the resonance. The narrower the resonance, \textit{i.e.} the smaller the full-width-at-half-maximum (FWHM), the more precisely the center of the resonance can be determined. The pronounced central peak, which is clearly distinguishable from the usually smaller side bands, marks the point of the maximal conversion and thus of the maximal radial energy and the shortest time of flight from the trap to the detector. Ramsey's method of time-separated oscillatory fields promises to lead to narrower central peaks compared to the conventional excitation scheme. Therefore, a variety of different Ramsey excitation schemes were investigated with the aim of finding the one best suited for precision mass spectrometry.
\newline
In \cite{Kre06} the line shapes are discussed in the quantum mechanical formalism as probability distributions for the partial conversion of a given initial state into a state of cyclotron motion. This general framework permits to take into account different assumptions on the initial state as well as statistical hypotheses, for example on phases. A more elementary
approach consists in calculating the time development of the radius of the cyclotron motion $R_{+}(t)$ from the given initial conditions, using Eq.~(\ref{alphaplus}),~(\ref{alphaminus}), and~(\ref{radius}).
\begin{figure}
\centering
{%
 \includegraphics[width=7cm,keepaspectratio]{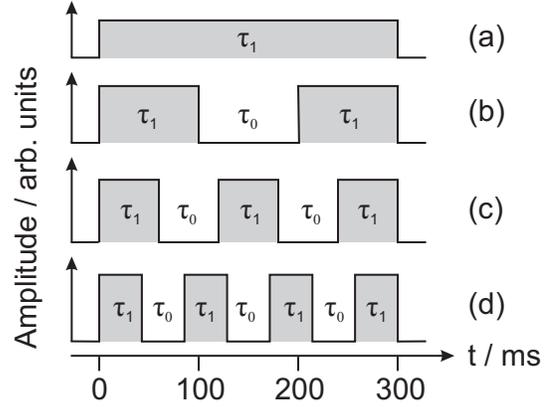}
        % !!! sonst 7 und 8cm
      } \vspace*{0.3cm} 
      \caption{Excitation schemes: (a) standard excitation, (b) excitation with two 100 ms
      Ramsey pulses, (c) excitation with three 60 ms Ramsey pulses, and (d) excitation with
      four 40 ms Ramsey pulses. The excitation amplitude is chosen in a way that at $\nu_c$ one
      full conversion from pure magnetron to pure cyclotron motion is obtained. Thus the sum of
      the grey colored areas is identical in all four schemes.} \label{Scheme}
\end{figure}
\subsubsection{The one-pulse excitation scheme}
In the conventional one-pulse excitation scheme the external driving field is applied for a certain time interval $\tau=\tau_{{\rm tot}}$ with constant amplitude, as shown in Fig. \ref{Scheme}. The lineshape represents the probability for the conversion of an initially pure state of magnetron motion into cyclotron motion as a function of the detuning $\delta$ of the quadrupole field. For given values of $\tau$ and $g$ it is obtained in the general formalism of~\cite{Kre06} as
\begin{equation}
F_{1}(\delta;\tau,g)
\;=\;\frac{4g^{2}}{\omega_{R}^{2}}\sin^{2}\left(\frac{\omega_{R}\,\tau}{2}\right)\;
. \label{1-pulse}
\end{equation}
Note that the maximum value 1 is reached when the driving field is at resonance, $\nu_{d}=\nu_{c}$, and the duration of the excitation is chosen
to satisfy the condition $g\tau=(2n+1)\pi/2$ with $n=0,1,2,$.... In different words, at resonance ($\delta=0$) complete conversion of a pure magnetron state into a state of pure cyclotron motion is achieved for excitation pulses of the duration of the \lq conversion time' $\tau_{c}=\pi/(2g)$  or odd multiples thereof.

This result can be verified by application of Eq.~(\ref{alphaplus}) and (\ref{radius}) with the initial condition $\alpha_{+}(0)=0$ and arbitrary $\alpha_{-}(0)$, and computing the time development of the radius of modified cyclotron motion:
\begin{equation}
R_{+}^{2}(\tau) \;=\;
\frac{4g^{2}}{\omega_{R}^{2}}\sin^{2}\left(\frac{\omega_{R}\,\tau}{2}\right)
\cdot R_{-}^{2}(0) \;=\; F_{1}(\delta;\tau,g) \cdot
R_{-}^{2}(0)\;.
\end{equation}
With respect to time the shape of the excitation pulse is rectangular. Thus, in frequency space it is expected that the excitation resembles the intensity ({\it i.e.} the modulus squared) of the Fourier transform of a rectangular profile, namely $(4g^{2}/\delta^{2})\sin^{2}(\delta\tau/2)$. The actual lineshape, however, differs in two important respects from this Fourier transform: (a) At resonance ($\delta=0$) it is a periodic function of $\tau$, describing the periodic conversion and reconversion of the magnetron and modified cyclotron modes, $F_{1}(\delta=0;\tau,g)=\sin^{2}(\pi\tau/\tau_{c})$, whereas the intensity of the Fourier transform increases proportionally to $\tau^{2}$. (b) The central peak is actually narrower than for the Fourier transform of a rectangle. This can be deduced from the position $\delta_{0}$ of the zero that separates the central peak from the first satellite peak, $(\delta_{0}\tau_c)^{2}=3\pi^2$ as compared to $(\delta_{0}\tau)^{2}=4\pi^{2}$ for the Fourier transform of the  rectangle, assuming $\tau=\tau_c$. 

\subsubsection{The two-pulse excitation scheme and more general schemes}
Although it is not obvious on first sight, a close formal analogy exists between nuclear magnetic resonance on the one hand and the interconversion of the magnetron and cyclotron motional modes of an ion in a Penning trap due to quadrupole excitation on the other hand. This was shown in \cite{Kre06} using the concept of a Bloch vector. It is therefore reasonable to expect that the use of Ramsey's method of separated oscillatory fields will lead to increased precision in mass spectrometry too.

A symmetric $n$-pulse Ramsey cycle of a total duration $\tau_{{\rm tot}}$ consists of $n$ excitation intervals of duration $\tau_{1}$ with ($n-1$) waiting intervals of duration $\tau_{0}$ in between, so that the total cycle time is $\tau_{{\rm tot}}=n\tau_{1}+(n-1)\tau_{0}$. Note that the $n$ excitation pulses of the rf-quadrupole field must be coherent in phase.

Assuming that the ion is initially in a state of pure magnetron motion, the probability for the conversion of magnetron quanta into quanta of the cyclotron motion by a 2-pulse Ramsey cycle with detuning $\delta=\omega_{d}-\omega_{c}$ has been calculated to be
\begin{eqnarray}
F_{2}(\delta;\tau_{0},\tau_{1},g) &=&
\frac{4g^{2}}{\omega_{R}^{2}} \left\{
\cos\left(\frac{\delta\tau_{0}}{2}\right)\sin\left(\omega_{R}\tau_{1}\right)\right.
\nonumber\\ & & \left. +\frac{\delta}{\omega_{R}}
\sin\left(\frac{\delta\tau_{0}}{2}\right)\right.\nonumber\\
& &\left.\left[\cos\left(\omega_{R}\tau_{1}\right)-1\right]
\right\}^{2}. \label{2-pulse}
\end{eqnarray}
This result and analogous ones for Ramsey excitation cycles with 3, 4, and 5 pulses can be found in \cite{Kre06}. If the frequency of the quadrupole field equals the cyclotron frequency $\nu_{c}$ and the amplitude of the field is chosen such that the $n$ excitation intervals exactly add up to the conversion time $\tau_{c}$, {\it i.e.} if the coupling constant $g$ satisfies the relation $n\tau_{1}=\tau_{c}=\pi/2g$, then the profile function (\ref{2-pulse}) reaches the value 1 at resonance, $F_{2}(\delta=0;\tau_{0},\tau_{c}/2,g)=1$.

An elementary derivation of the 2-pulse profile function $F_{2}$ in Eq.~(\ref{2-pulse}) is possible by applying Eq.~(\ref{alphaplus}), (\ref{alphaminus}), and (\ref{radius}) three times successively to the time intervals $0\leq t\leq \tau_{1}$, $\tau_{1}\leq t\leq \tau_{1}+\tau_{0}$, and $\tau_{1}+\tau_{0}\leq t\leq 2\tau_{1}+\tau_{0}=\tau_{{\rm tot}}$,    with the initial condition $\alpha_{+}(0)=0$ and arbitrary $\alpha_{-}(0)$, in order to compute the endpoint of the time development of the radius of the cyclotron motion, $R_{+}(\tau_{{\rm tot}})$. The first calculation yields the initial values $\alpha_{\pm}(\tau_{1})$ for the waiting period, the second calculation with $g=0$ yields the phase change during the waiting
period and thus the initial values $\alpha_{\pm}(\tau_{1}+\tau_{0})$ for the second excitation
period, the third step finally results in $\alpha_{\pm}(2\tau_{1}+\tau_{0})=\alpha_{\pm}(\tau_{{\rm tot}})$ and thus in a result for the final radius of the  cyclotron motion $R_{+}(\tau_{{\rm tot}})$. Note that $\chi_{d}$ in Eq.~(\ref{alphaplus}) and (\ref{alphaminus}) has to be replaced in the second and third step by the corresponding phases of the quadrupole field, $\omega_{d}\tau_{1}+\chi_{d}$ and $\omega_{d}(\tau_{0}+\tau_{1})+\chi_{d}$, respectively. The final result is
\begin{eqnarray}
R_{+}^{2}(\tau_{{\rm tot}}) & = & F_2(\delta;\tau_0,\tau_1,g)\cdot R_{-}^{2}(0).   \label{radius-2-pulse}
\end{eqnarray}
The time development of the ion orbit during the two excitation periods is shown in Fig.~\ref{ConversionMaCy}a and Fig.~\ref{ConversionMaCy}b, during the waiting time the ion follows a rosette shaped orbit.

\begin{figure}
\resizebox{0.5\textwidth}{!}
{%
 \includegraphics{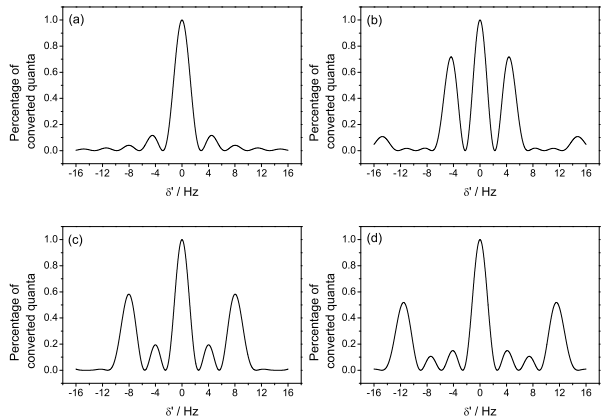}
        % !!! sonst 7 und 8cm
       } \vspace*{0.3cm} 
       \caption{Percentage of converted quanta in the case of a quadrupolar
        excitation near $\omega_c$ as a function of the frequency detuning $\delta '=\delta/2\pi$ for different excitation schemes shown in
        Fig.~\ref{Scheme}. The total cycle time of
        all schemes is 300 ms. The line profile of a one-pulse excitation with 300 ms
        excitation time is shown in (a). (b) represents the line profile of a two-pulse
        excitation, each of 100 ms duration. The excitation time as well as the waiting time of
        the three-pulses excitation scheme in (c) is 60 ms. In (d) the four 45 ms pulses are
        interrupted by 40 ms waiting periods. The center frequency ($\delta=0$) is the cyclotron frequency
        $\omega_c$ for a given ion mass $m$.}
    \label{Conversion1-4}
\end{figure}

Plotting the profile function (\ref{2-pulse}) with a fixed value of the waiting time $\tau_{0}$ as a function of the detuning $\delta$ one obtains the spectral lineshape of the 2-pulse Ramsey cycle. It bears a resemblance to the Fourier transform of a signal consisting of two rectangular pulses, but as for a single pulse there are important and characteristic differences. Lineshapes for higher order Ramsey cycles are calculated in an analogous fashion using the results of \cite{Kre06}. In Fig. \ref{Conversion1-4} generic results are displayed for $n=1,2,3,4$, assuming a total cycle time $\tau_{{\rm tot}}=300$ ms and $\tau_{1}=\tau_{c}/n$ for all excitation schemes. Note that with the $n$-pulse excitation scheme a spectral distribution is obtained in which the valley between the first major sideband and the central peak contains ($n-2$) small peaks, while the distance between the first major sideband and the central peak increases with $n$.

\section{Implementation at the mass spectrometer ISOLTRAP}
\label{setup}
\subsection{Experimental setup}
\begin{figure}
\resizebox{0.5\textwidth}{!}
{%
 \includegraphics{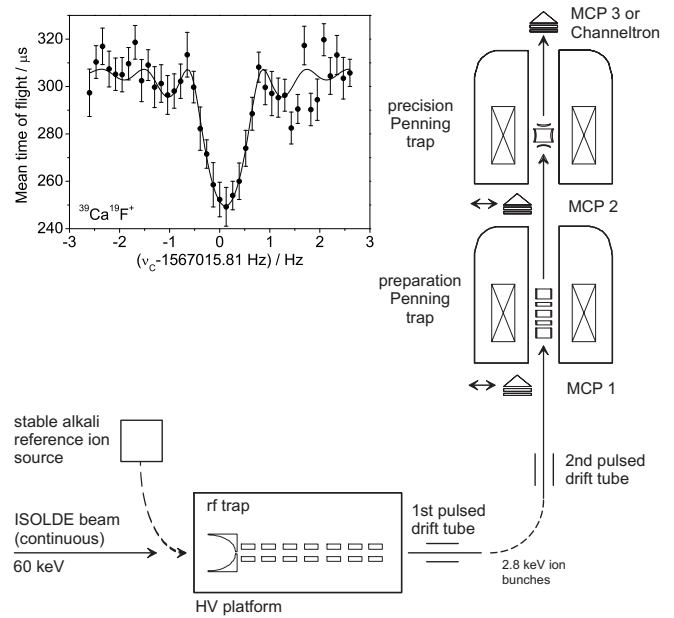}
} \vspace*{0.3cm} 
    \caption{Schematic drawing of the mass spectrometer ISOLTRAP including the RFQ trap, the
    preparation and precision Penning traps, as well as the reference ion source. In the inset
    a resonance for $^{39}\mbox{Ca}^{19}\mbox{F}^+$ ions with an excitation period of 900 ms is
    shown. To monitor the ion transfer and to record the time-of-flight resonance for the 
    determination of the cyclotron frequency, micro-channel-plate (MCP) detectors or a
    channeltron are used.} \label{fig:ISOLTRAPSetup}
\end{figure}
The triple-trap mass spectrometer ISOLTRAP~\cite{Bla05,Bol96} installed at the on-line  facility ISOLDE/CERN~\cite{Kug00,Äys04} is dedicated to high-precision mass measurements of radioactive nuclides. It reaches a relative mass uncertainty of below $10^{-8}$~\cite{Kel03}.
A Paul trap and two Penning traps are the main parts of the apparatus as shown in Fig.~\ref{fig:ISOLTRAPSetup}. The ions from the continuous 60-keV radioactive beam of ISOLDE or from the stable alkali ion source are captured in-flight by a gas-filled linear Paul trap with segmented rods~\cite{Her01a}. Here, the ions are accumulated, cooled, and bunched. From the Paul trap the ions are guided, after passing two pulsed drift tubes in order to reduce the potential energy from 60 keV to about 100 eV, to a buffer-gas filled cylindrical Penning trap~\cite{Rai97}. This preparation trap is located in a superconducting magnet of 4.7-T field strength. Here, the ions are mass-selectively cooled~\cite{Sav91} with a resolving power of up to $10^5$. Thus, an isobarically clean ion cloud is obtained, which is transferred to the precision Penning trap. This hyperbolic Penning trap ~\cite{Bol96}, which is placed in a second superconducting magnet of 5.9-T field strength with a field homogeneity of $10^{-7}$ to $10^{-8}$ within one cubic centimeter and a temporal stability of $\frac{\delta B}{B}\frac{1}{\delta t}\approx 2\cdot 10^{-9}$/h, provides the possibility to resolve even low-lying excited nuclear states~\cite{Van04,Bla04a} and serves for the actual mass measurement of nuclides with half-lives even below 100 ms~\cite{Bla03c,Kel04}.
\newline
In order to manipulate the ion motion in the precision trap by the application of external rf-fields, the ring electrode of the trap is four-fold segmented, as illustrated in Fig.~\ref{Dipol_Quadrupol}. First, the magnetron radius of all ions is increased via a dipolar excitation at the magnetron frequency~\cite{Bla03b}. If required, there is the possibility to remove unwanted contaminations via a mass-selective dipolar excitation at the corresponding modified cyclotron frequency. Thereby the ions' cyclotron radius is increased until the ions hit the electrode and are lost. A quadrupolar excitation at $\nu_c$ converts the magnetron motion into cyclotron motion. Finally, the ions are ejected out of the trap and pass the gradient of the magnetic field, which interacts with the orbital magnetic moment of the ions and accelerates them in the axial direction to the detector. This acceleration is proportional to the strength of the ions' magnetic moment, \emph{i.e.} to the radial energy obtained by the excitation. The time of flight after ejection from the trap to the detector is measured. This procedure is repeated for different frequencies of the quadrupolar excitation in the precision trap around the expected value of the cyclotron frequency. By the determination of the mean time of flight for the different excitation frequencies a time-of-flight resonance curve is recorded (see inset of Fig.~\ref{fig:ISOLTRAPSetup}). For appropriate excitation parameters the minimum time of flight is measured at the cyclotron frequency~\cite{Koe95}. For reference measurements, \emph{i.e.} to calibrate the magnetic field, ions with well-known mass from a stable alkali ion source are used.
\newline
The experimental standard deviation $\sigma(\nu_c)$ of the cyclotron frequency $\nu_c$ is a
function of the resolving power of the precision trap, \emph{i.e.} the quadrupolar excitation time $T_q$, and the total number $N_{tot}$ of recorded ions. The resolving power is Fourier limited by the duration of the quadrupolar excitation, which itself is limited by the half-life of the ion of interest in case of short-lived radionuclides. An empirical formula \cite{Bol01}
describes this relation:
\begin{eqnarray}
 \frac{\sigma(\nu_c)}{\nu_c}=\frac{1}{\nu_c}\frac{c}{\sqrt{N_{tot}}\cdot T_q}\mbox{
 ,}
\end{eqnarray}
where $c$ is a dimensionless constant. In a large number of measurements with carbon clusters the constant $c$ was determined for the ISOLTRAP mass spectrometer to be $c=0.898(8)$~\cite{Kel03}.

\subsection{Reduction of the line-width}
\begin{figure}
\resizebox{0.5\textwidth}{!}
{%
 \includegraphics{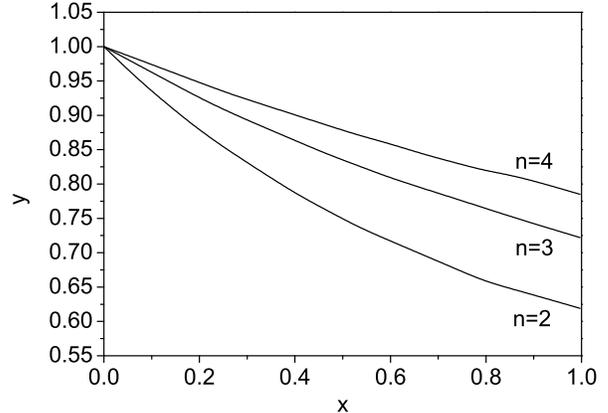}
        % !!! sonst 7 und 8cm
   } \vspace*{0.3cm} 
   \caption{Reduction of the width of the central peak due to the Ramsey method for constant
    cycle time $\tau_{{\rm tot}}$. For symmetrical $n$-pulse excitation a Ramsey cycle is
    composed of $n$ excitation pulses of duration $\tau_{1}=\tau_{c}/n$, where $\tau_{c}$ is
    the conversion time, separated by ($n-1$) waiting periods of duration $\tau_{0}$. Thus the
    total cycle time is $\tau_{{\rm tot}}=\tau_{c}+(n-1)\tau_{0}$. The variable 
    $y=|\delta_{0}^{(n)}\tau_{{\rm tot}}/(\pi\sqrt{3})|$ is a measure for the FWHM of the
    central peak for $n$-pulse excitation relative to the corresponding quantity for 1-pulse
    excitation, $x=(n-1)\tau_{0}/\tau_{{\rm tot}}$ is the fraction of the total cycle time
    spent during waiting periods.} \label{widthreduction}
\end{figure}

Having discussed the standard quadrupolar excitation, what is the advantage of Ramsey's method of time-separated oscillatory fields? In many experimental situations, the total time available for a complete measurement cycle has an upper limit, for example due to the lifetime of the
radioactive species under investigation. Thus, a precision gain simply by increasing the waiting time $\tau_{0}$ may not be feasible due to experimental limitations. Therefore, different excitation schemes are compared with respect to the predicted width of the central peak, assuming that a total time $\tau_{{\rm tot}}$ is available to perform one complete Ramsey measuring cycle. For a symmetric $n$-pulse excitation it is $\tau_{{\rm tot}}=n\tau_{1}+(n-1)\tau_{0}$, where $\tau_{1}$ denotes the duration of an excitation interval and $\tau_{0}$ the duration of a waiting interval. For $\tau_{1}=\tau_{c}/n$ and an excitation at resonance a complete conversion of a pure magnetron state into a pure state of cyclotron motion occurs, \textit{i.e.} the highest possible degree of conversion.  A suitable parameter to describe the width of the central peak is the position $\delta_{0}^{(n)}$ of the zero separating the central peak from the first side band. For a one-pulse excitation $(\delta_{0}^{(1)}\tau_{{\rm tot}})^{2}=3\pi^{2}$. Therefore, for an $n$-pulse excitation the
variable $y=|\delta_{0}^{(n)}\tau_{{\rm tot}}/(\pi\sqrt{3})|=|\delta_{0}^{(n)}/\delta_{0}^{(1)}|$ represents the ratio of the zeros for $n$-pulse and 1-pulse excitation. To a very good approximation this variable equals the
ratio of the full-width-half-maximum of the two excitation schemes. In Fig. \ref{widthreduction} these ratios have been plotted for $n=2,3,4$ as a function of
$x=(n-1)\tau_{0}/\tau_{{\rm tot}}$, {\it i.e.} the percentage of the total cycle time $\tau_{{\rm tot}}$ spent during waiting periods. From the graph it is obvious that with 60\% the two-pulse excitation scheme offers the largest width reduction relative to a one-pulse excitation. This excitation scheme is therefore favored for the application in high-precision mass spectrometry.
\subsection{Time-of-flight detection technique at ISOLTRAP}
\begin{figure}[h!]
    \centering
    \begin{minipage}[c]{.5\textwidth}
        \includegraphics[width=8cm]{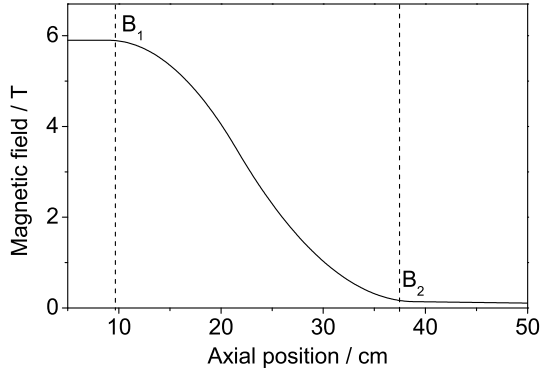}
    \end{minipage}
     \caption{The magnetic field amplitude from the precision trap to the end of the conversion
     section. Between the two marked points the amplitude decreases about 90\% of its origin
     value.}
         \label{mafi}
\end{figure}
As described above, the frequency determination via a time-of-flight detection technique~\cite{Gra80} is based on the interaction of the magnetic moment of the orbiting ion with the magnetic field gradient (Fig.~\ref{mafi}). Thus, for a detailed comparison with experimental data it is necessary to convert the theoretical line shapes for the energy conversion into time-of-flight spectra, which will be discussed in the following. The kinetic  energy in the radial motions is predominantly due to the cyclotron motion; the contribution of the magnetron mode is negligible because of $\omega_-\ll\omega_+$: 
\begin{eqnarray}
E_{r}(\omega_d) & = & E_{r}^{kin}(\omega_d)+E_{r}^{pot}(\omega_d)\nonumber\\
& = & \frac{1}{2}m\cdot\left(R_+^2(\tau_{tot},\omega_d)\omega_+^2+R_-^2(\tau_{tot},\omega_d)\omega_-^2\right)\nonumber\\
& & -\frac{1}{2}m\cdot\omega_+\omega_-\left(R_+^2(\tau_{tot},\omega_d)+R_-^2(\tau_{tot},\omega_d)\right)\nonumber\\
& \approx & \frac{1}{2}m\cdot R_+^2(\tau_{tot},\omega_d)\omega_+^2
{ .}
\end{eqnarray} 
Here, $R_{\pm}(\tau_{tot},\omega_d)$ are the radii of the two radial modes after the quadrupolar excitation has been applied with the frequency $\omega_d$. The magnetic moment of an ion with kinetic energy $E_r(\omega_d)$ in a magnetic field $\vec{B}=B\cdot\vec{e}_z$ can be written as $\vec{\mu}(\omega_d)=\left[E_r(\omega_d)/B\right]\vec{e}_z$. The interaction with the gradient of the magnetic field causes an axial force $\vec{F}_z(\omega_d)=-\vec{\mu}(\omega_d)\cdot\nabla \vec{B}_z$ on the ion, which leads to a reduction of the time of flight from the trap to the detector. This time of flight can be calculated with~\cite{Koe95}
\begin{eqnarray}
T(\omega_d)=\int_{z_0}^{z_1}\left\{\frac{m}{2\left[E_0-q\cdot V(z)-\mu (\omega_d)\cdot B(z)\right]}\right\}^{\frac{1}{2}}dz \mbox{ ,}\nonumber\\
\end{eqnarray}
where $E_0$ is the total initial energy of the ion, \textit{V(z)} and \textit{B(z)} are the electric and magnetic fields, respectively, along the way from the trap to the detector. At $\omega_d=\omega_c$ the magnetic moment is maximal and thus the time of flight minimal.
\begin{figure}
\resizebox{0.5\textwidth}{!}
{%
 \includegraphics{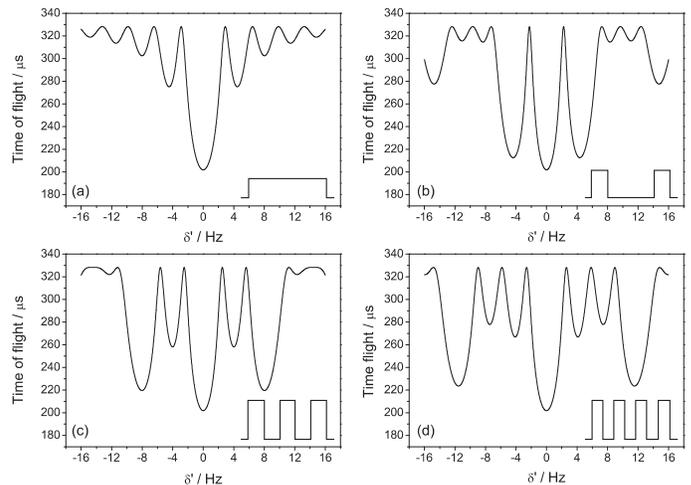}
        % !!! sonst 7 und 8cm
   } \vspace*{0.3cm}
    \caption
    {Calculated time-of-flight cyclotron resonances for different excitation schemes. The total
     excitation and waiting time in the precision trap is 300 ms. The time of flight of a
     one-pulse excitation with 300 ms duration is shown in (a). (b) shows the time of flight of
     a two-pulse excitation, each of the pulses being 100 ms long. The three pulse excitation
    (c) is done by three 60-ms excitation periods and two 60-ms waiting periods. In (d) the
     four 45-ms pulses are interrupted by 40-ms waiting periods.}
    \label{TOF1-4}
\end{figure}
Typical theoretical time-of-flight cyclotron-resonance curves for different excitation schemes using the radial energies calculated from the equations of conversion (see Eq.~(\ref{1-pulse})
and~(\ref{2-pulse}) and Ref.~\cite{Kre06}) are shown in Fig.~\ref{TOF1-4}. In each graph the time of flight of the ions from the trap to the detector is plotted as a function of the frequency detuning $\delta '=\delta/2\pi$ with respect to the cyclotron frequency $\nu_c$. 
\section{Results}
 \label{results}
 \begin{figure}
\resizebox{0.5\textwidth}{!}
{%
 \includegraphics{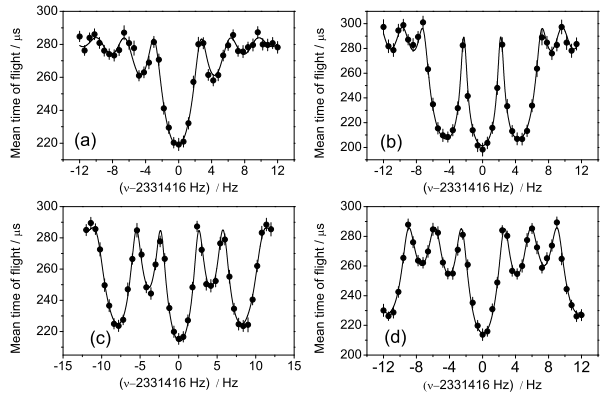}
        % !!! sonst 7 und 8cm
   } \vspace*{0.3cm}
    \caption{The measured mean time of flight as a function of the quadrupolar excitation
    frequency with predicted curves fitted to the data. Here $^{39}\mbox{K}^+$ ions from the stable
    alkali ion source were used. (a) Resonance of the conventional one-pulse excitation scheme with 300 ms excitation time,
    (b) a two-pulse excitation scheme with two times 100 ms excitation and 100 ms waiting time.
    (c) three pulses, each 60 ms, interrupted by two waiting periods of 60 ms, (d) four
    45 ms excitation pulses and three waiting periods of 40 ms.}
    \label{ResonanceCurves}
\end{figure}
\begin{figure}
\resizebox{0.5\textwidth}{!}
{%
 \includegraphics{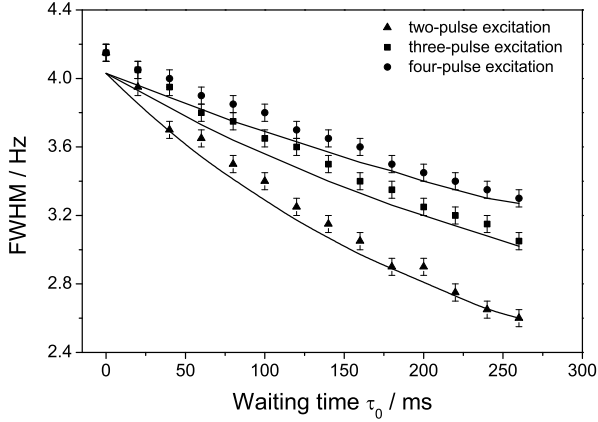}
        % !!! sonst 7 und 8cm
   } \vspace*{0.3cm}
    \caption{The full-width-half-maximum values of the time-of-flight cyclotron resonances
    as a function of the sum of all waiting periods $\tau_0=\sum_i\tau_0^i$ for two, three and four excitation pulses. The total cycle time is $\tau_{\rm tot}$=300 ms. The
    solid lines are the theoretically calculated FWHM values.}
    \label{FWHM_Ex_2_3_4}
\end{figure}
\begin{figure}
\resizebox{0.5\textwidth}{!}
{%
 \includegraphics{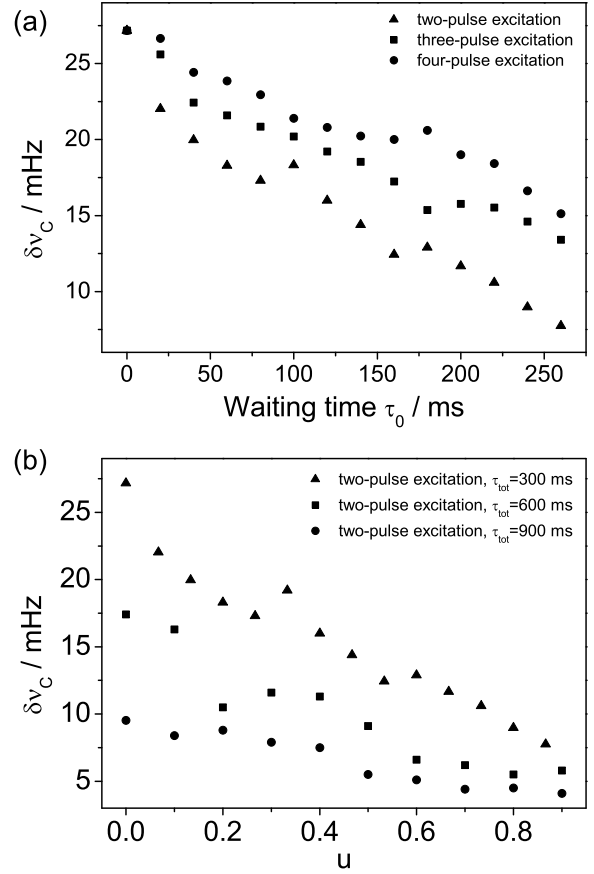}
        % !!! sonst 7 und 8cm
    } \vspace*{0.3cm}
    \caption{In (a) the uncertainty of the measured cyclotron frequency is given as a
    function of the waiting time for the two-pulse, three-pulse, and four-pulse excitation
    scheme. In (b) the uncertainty of the measured cyclotron frequency is given for the
    two-pulse scheme with a total cycle of 300 ms, 600 ms, and 900 ms, where
    $u=(n-1)\tau_{0}/\tau_{{\rm tot}}$ is the fraction of the total cycle time spent during
    waiting periods.}
    \label{Uncertainty}
\end{figure}

To confirm the calculations, \emph{i.e.} to determine the line-width reduction and, most importantly, to specify the precision gain due to the Ramsey excitation method, more than 300
time-of-flight resonance curves with different excitation schemes were recorded with the Penning trap mass spectrometer ISOLTRAP. The ion species for all measurements was $^{39}\mbox{K}^+$ provided by the stable alkali off-line ion source. Each resonance consists of about 2500 ions in order to have identical statistics. To minimize ion-ion interactions only time of flight measurements with at most five ions in the trap were taken into account. The efforts were concentrated on the specification of the uncertainty in the frequency determination for different excitation schemes. Figure \ref{ResonanceCurves} shows time-of-flight cyclotron resonances for the one-, two-, three-, and four-pulse excitation scheme.
A fit of the theoretically expected line shape (solid line) to the data points allows the determination of the FWHM and the cyclotron frequency $\nu_c$ along with its uncertainty $\delta\nu_c$. To perform these fits the standard evaluation program of ISOLTRAP \cite{Kel03} was extended in order to analyze the measured cyclotron resonances using the Ramsey method.
\begin{figure}
\resizebox{0.5\textwidth}{!}
{%
 \includegraphics{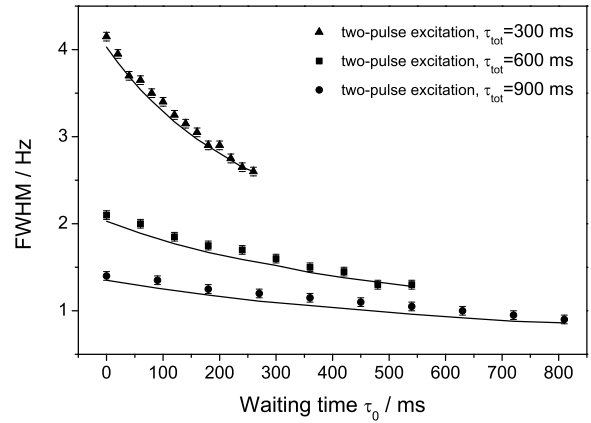}
        % !!! sonst 7 und 8cm
   } \vspace*{0.3cm}
    \caption{The full-width-at-half-maximum (FWHM) values
    of the time-of-flight cyclotron resonances as a function of the waiting time
    for the two-pulse excitation scheme is given. The total cycle times are $\tau_{\rm tot}=300$ ms
    (squares), 600 ms (circles), and 900 ms (triangles).
    The bold solid lines are the theoretically calculated
    values. Since the determination of the FWHM was performed manually
     from the fit curves, the error bars
    are conservatively estimated to be $\pm 0.05$ Hz.}
    \label{FWHM_Ex_3_6_9}
\end{figure}
\newline
The fit results concerning the FWHM are presented in Fig.~\ref{FWHM_Ex_3_6_9} and Fig.~\ref{FWHM_Ex_2_3_4}. Figure~\ref{FWHM_Ex_3_6_9} shows results obtained with a two-pulse
excitation scheme of different overall cycle times ($\tau_{\rm tot}=$ 300 ms, 600 ms, 900 ms). The FWHM is given as a function of the waiting period. Due to field inhomogeneities
    and ion-ion interactions the data points are shifted slightly to higher FWHM values compared to theory. Similar results are shown in Fig.~\ref{FWHM_Ex_2_3_4} for different numbers of excitation pulses, where the total cycle time in the precision trap is constant $\tau_{\rm tot}=300$ ms. The experimental values are on average 0.1 Hz higher than the theoretical ones. This line-broadening effect is due to the electric and magnetic field imperfections and ion-ion interactions, which were not taken into account in the calculations described above. A significant reduction of the FWHM for shorter excitation pulses with longer waiting periods in between can be observed.
\newline
In Tab.~\ref{tabular_FWHM} the experimental results are summarized. ``cycle time'' is the duration of the total cycle. The maximum FWHM is the one of the standard one-pulse resonance curve. The minimum FWHM is measured using the longest possible waiting time, \emph{i.e.} shortest possible excitation time, which is determined by the maximal possible amplitude of the quadrupolar excitation field required to obtain one full conversion from pure magnetron
to pure cyclotron motion. The last column of Tab.~\ref{tabular_FWHM} gives the maximum line-width reduction (reduction gain=\newline
(max. FWHM - min. FWHM)/max. FWHM) using the different Ramsey excitation schemes. A remarkable reduction of close to 40\% of the normal line-width is observed, similar to the results in the original work of Ramsey~\cite{Ram90}. The reduction in line-width is especially important in context with the achievable resolving power
$R=m/\Delta m=\nu/\Delta \nu$. As theoretically predicted, the largest possible reduction is obtained by a two-pulse excitation scheme. However, the relative gain in reduction depends only weakly on the total cycle time $\tau_{tot}$ (see Tab.~\ref{tabular_FWHM} and Fig.~\ref{FWHM_Ex_3_6_9}).
\newline
\begin{table}
  \caption
   {The maximum and minimum experimental FWHM of the different excitation schemes for different
   cycle times are given. In addition the reduction gain is calculated. For further explanation
   see text.}\label{tabular_FWHM}
  \begin{center}
   \begin{tabular}{|c|c|c|c|c|}
        \hline
        number & cycle time & max. FWHM & min. FWHM & reduction \\
        of pulses & $\tau_{tot}$ & Hz & Hz & gain $\%$ \\\hline
        2 & 300 & 4.1 (0.1) & 2.6 (0.1) & 37.3 (2.0)\\\hline
        2 & 600 & 2.1 (0.1) & 1.3 (0.1) & 38.1 (4.0)\\\hline
        2 & 900 & 1.4 (0.1) & 0.9 (0.1) & 35.7 (5.9)\\\hline
        3 & 300 & 4.1 (0.1) & 3.0 (0.1) & 26.5 (2.2) \\\hline
        4 & 300 & 4.1 (0.1) & 3.3 (0.1) & 20.5 (2.7) \\
        \hline
   \end{tabular}
   \end{center}
\end{table}
\hspace{-6pt} 
In Fig.~\ref{Uncertainty} the experimental uncertainty $\delta\nu_c$ of the measured cyclotron frequencies for different numbers of pulses and different length of the total cycle time is plotted versus the waiting time $\tau_0$. Each data point represents the mean value of three to ten individual measurements. The uncertainty $\delta\nu_c$ decreases with increasing waiting time. This is expected due to the decreasing FWHM at longer waiting times. If the cyclotron frequency uncertainty would only depend on the FWHM, a similar behavior as observed in Fig.~\ref{FWHM_Ex_2_3_4} would be expected. However, as mentioned before, it is obvious that there is also an effect of the overall line-shape, especially of the steepness of the curve and the pronounced sidebands, on the uncertainty in the frequency determination of $\nu_c$.
\newline
The excitation scheme used for the data points given in Fig.~\ref{Uncertainty} (a) consists of two, three, and four pulses, where the total excitation cycle is fixed to 300 ms. The uncertainty is obviously decreasing for shorter excitation pulses, \emph{i.e.} longer waiting times $\tau_0$. In case of the two-pulse excitation it can be reduced by more than a factor of three (from $\delta\nu_c\approx27$ mHz down to $\delta\nu_c\approx8$ mHz) as compared to the conventional procedure just by changing the excitation scheme to the two-pulse method. The uncertainty using the three- and four-pulse excitation scheme decreases to $\delta\nu_c\approx13$ mHz and $\delta\nu_c\approx15$ mHz, respectively. 

The result is summarized in Tab.~\ref{tabular_uncertainty}, where the maximal and the minimal uncertainty for all investigated excitation schemes under identical experimental conditions are listed. As for the investigation of the FWHM, the scheme where the largest reduction in the cyclotron frequency uncertainty can be achieved is the two-pulse Ramsey scheme. Comparing the two-pulse excitation scheme with 300 ms, 600 ms, and 900 ms total cycle time, the tendency is similar to the results obtained for the FWHM of the resonances (see Fig.~\ref{Uncertainty} (b)). The slightly decreasing gain factor for two-pulse schemes with longer total cycle times must be assigned to the relative decrease of the statistical uncertainty in comparison to the constant systematic uncertainty. 
Unsymmetric excitation schemes where the individual Ramsey pulses have different lengths have also been investigated. However, the symmetric two-pulse
excitation scheme remains the best one in respect to line-width reduction and uncertainty gain (see Tab.~\ref{tabular_uncertainty}).
\begin{table}
  \caption{The maximum and the minimum experimental uncertainties of different excitation
  schemes and excitation times are listed. The improvement factor is given in the last column.}
   \label{tabular_uncertainty}
  \begin{center}
   \begin{tabular}{|c|c|c|c|c|}
        \hline
        number & cycle time & max. & min. & Improvement  \\
        of pulses & $\tau_{tot}$ & uncertainty & uncertainty & factor  \\
        & &  Hz & Hz &  \\\hline
        2 & 300 & 0.027 & 0.008 & 3.4 (0.4) \\\hline
        2 & 600 & 0.017 & 0.006 & 2.8 (0.3) \\\hline
        2 & 900 & 0.010 & 0.004 & 2.5 (0.3) \\\hline
        3 & 300 & 0.027 & 0.013 & 2.1 (0.2) \\\hline
        4 & 300 & 0.027 & 0.015 & 1.8 (0.2) \\
        \hline
   \end{tabular}
   \end{center}
\end{table}

\section{First online mass spectrometry application of the Ramsey method}
\begin{figure}
\resizebox{0.5\textwidth}{!}
{%
 \includegraphics{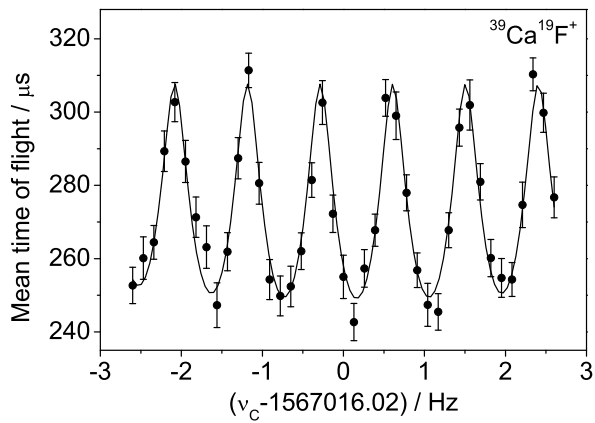}
        % !!! sonst 7 und 8cm
    } \vspace*{0.3cm}
    \caption{Time of flight for $^{39}\mbox{Ca}^{19}\mbox{F}^+$. A two-pulse Ramsey scheme was chosen with two 100
    ms duration excitation periods interrupted by a 700 ms waiting period. The solid curve is a
    fit of the theoretically expected line shape to the data.}
    \label{OnlineResonances}
\end{figure}
The first online mass measurement by the Ramsey excitation method was carried out for the short-lived nuclides $^{38}\mbox{Ca}$ and $^{39}\mbox{Ca}$, which have half-lives of only $T_{1/2}(^{38}\mbox{Ca})=440$ ms and $T_{1/2}(^{39}\mbox{Ca})=860$ ms, respectively~\cite{AME03}. In order to suppress $^{38}\mbox{K}^+$ contaminations, the $^{38}\mbox{Ca}^+$ ions were delivered in form of the sideband molecule $^{38}\mbox{Ca}^{19}\mbox{F}^+$. In the inset of Fig.~\ref{fig:ISOLTRAPSetup} a resonance curve of $^{39}\mbox{Ca}^{19}\mbox{F}^+$ is shown for which the ions were exposed to a continous quadrupolar radiofrequency excitation of 900 ms duration. The cyclotron frequency of this resonance has been determined with an uncertainty of $\delta\nu_c=32$ mHz. Fig.~\ref{OnlineResonances} shows a resonance of the same species with the same number of collected ions ($\approx 2500$), where a two-pulse Ramsey excitation scheme was used. The two
excitation pulses had a duration of 100 ms interrupted by a waiting period of 700 ms. Thus, the total time $\tau_{{\rm tot}}=900$ ms for which the ions remained in the trap was identical. Here, the cyclotron frequency was determined with an uncertainty of only $\delta\nu_c=9$ mHz. In comparison, the statistical error in the frequency determination could be reduced by more than a factor of three, keeping the number of ions and the storage time constant. This is a tremendous gain factor, especially for mass measurements on short-lived radionuclides since the required measurement time to reach a certain statistical uncertainty can be reduced by almost an order of magnitude.

\section{Conclusions and outlook}
The experimental studies described in this paper demonstrate that Ramsey's method of time-separated oscillatory fields
can be applied to excite the ion motion in a Penning trap. The Ramsey technique improves significantly the statistical uncertainty in high-precision mass
spectrometry on short-lived radionuclides. We performed 
systematic experimental investigations for different excitation patterns with two, three, and four excitation pulses. We observed the expected reduction of the line width of almost a factor of two along with a gain in precision in the frequency determination. This leads to an important gain in precision of the frequency determination. The new findings were demonstrated  for a stable nuclide as well as for a short-lived radionuclide in an on-line measurement. 

An optimized Ramsey excitation scheme with two pulses of short duration interrupted by a long waiting period results in an
improvement of the statistical uncertainty in the cyclotron-frequency determination by more
than a factor of three compared to the conventional scheme, without any further experimental changes as \textit{e.g.} the number of
detected ions for a resonance or reduction of scan detuning width. Since the Ramsey method opens a door to higher precision in Penning trap mass spectrometry, the application in other Penning trap setups is already in preparation and under investigation, as e.g. at SMILETRAP (Stockholm) using highly-charged stable ions~\cite{Ber02} or at SHIPTRAP (GSI, Darmstadt) using short-lived fission fragments~\cite{Rah06}. It can be expected that the Ramsey method will find a wide-spread application in high-precision mass spectrometry of atomic nuclei.

\end{document}